\documentclass[12pt]{article}
\usepackage{amssymb}
\usepackage{amsmath}
\usepackage{epsfig}
\begin{document}

\title{Deterministic Bak-Sneppen model. Lyapunov spectrum and avalanches as return
times}
\author{R. Vilela Mendes\thanks{%
Grupo de F\'{\i }sica Matem\'{a}tica, Complexo Interdisciplinar,
Universidade de Lisboa, Av. Gama Pinto, 2 - 1649-003 Lisboa, Portugal} 
\thanks{%
Laborat\'{o}rio de Mecatr\'{o}nica, DEEC, Instituto Superior T\'{e}cnico,
Av. Rovisco Pais, 1049-001 Lisboa, Portugal} \thanks{%
e-mail: vilela@cii.fc.ul.pt}}
\date{}
\maketitle

\begin{abstract}
A deterministic version of the Bak-Sneppen model is studied. The role of the
Lyapunov spectrum in the onset of scale-free behavior is established and
avalanches are interpreted as return times to a zero-measure set. The
problem of accurate determination of the scaling exponents near the critical
barrier is addressed using a characteristic function approach. A general
equation for return times to a small measure set is established and used to
obtain information on the nature of Bak-Sneppen dynamics.
\end{abstract}

\section{Introduction}

In Ref.\cite{Vilela1} several structure-generating mechanisms, in
multi-agent dynamical systems, have been analyzed. A {\it structure} in a
dynamical system is characterized by the emergence of phenomena at scales
different from those typical of the dynamics of the agents, when they evolve
in isolation. A {\it space-structure} corresponds to phenomena operating at
a scale much larger than the typical size of one agent and a {\it %
time-structure} is a phenomenon with a time scale much longer than the cycle
time of the agents.

An important role in the generation of time-structures is played by
modifications to the Lyapunov spectrum, occurring either as a result of the
agents' interaction or of a change of parameters. A new structure is created
each time one Lyapunov exponent crosses zero from above. In particular, in
one of the mechanisms described in \cite{Vilela1}, the joint effect of chaos
in the individual agent evolution law and extremal dynamics, implies that
all Lyapunov exponents tend to $0^{+}$ in the infinite agents limit. This
leads to a characteristic behavior free of time scales. It has been
conjectured in Ref.\cite{Vilela1} that it is this mechanism that is behind
many of the dynamical manifestations of what has been called self-organized
criticality (SOC). Here, this conjecture is pursued by analyzing a situation 
\cite{Head} where breakdown of self-organized criticality is exhibited. It
is shown that the breakdown of SOC may in fact be understood from the
behavior of the Lyapunov spectrum.

There is some controversy concerning a rigorous definition of SOC. Here only
the absence of natural time scales is emphasized, without much concern about
space scales, nature of the driving, separation of time scales and other
relevant issues useful for a precise characterization of SOC.

Absence of time scales, as a property of dynamics, is naturally related to
the Lyapunov spectrum. Time scales disappear whenever the Lyapunov exponents
vanish. This leads in a natural way to power laws for both time and space
correlation. However, scaling of the avalanches, that is, a power law for
the return times to the self-organized state, is a subtler effect. This is
easy to understand when the multi-agent system is formulated as a
measure-preserving dynamical system. If the self-organized state $A$, which
serves as reference for the counting of return times $\left( k\right) $, is
a non-zero measure set $\left( \mu \left( A\right) \neq 0\right) $, the
measure $\mu \left( A\right) $ itself serves as a natural time scale. Then,
the large time behavior of the return times distribution would be dominated
by an exponential factor $\exp \left( -k\nu \left( \mu \right) \right) $.
Therefore, for cases that fit in the ergodic dynamical systems setting,
power laws may occur only if the reference set has vanishing measure. In the 
$\mu \left( A\right) \rightarrow 0$ limit the return times distribution is
dominated by the pre-factor that multiplies the exponential and this one
might be or not be a power law.

When $\mu $ is an ergodic measure, by Kac's lemma, the mean return time to a
set $A$ is $1/\mu \left( A\right) $. Therefore when $\mu \left( A\right) =0$%
, the numerical evaluation of the return time (avalanche) law has to be
carried out for a set slightly larger than $A$. This implies that it may be
difficult to disentangle the prefactor dependence from the exponential one,
leading to some uncertainty about the exact values of the numerically
measured scaling exponents. In particular because, as seen in Sect. 2, the
exponent factor $\nu \left( \mu \right) $ may be a non-trivial function of
the measure. Instead of obtaining the return times distribution from the
histogram, a more robust way is described in Sect.2, based on the
construction of the characteristic function.

Treating the multi-agent system as a deterministic measure-preserving
dynamical system, tools from ergodic theory may be applied. An important
issue is the characterization of asymptotic sets and invariant measures as
well as the nature of the relevant attractors and repellers of the
scale-free system. These questions are discussed for a deterministic version
of the Bak-Sneppen model\cite{Bak1}. The nature of the self-organized state
as a zero-measure subset of the invariant measure is put into evidence.
Then, a probabilistic avalanche equation, written for the return times, is
used to extract the non-trivial nature of the avalanche process from
simulation data.

\section{The deterministic Bak-Sneppen model}

The original Bak-Sneppen model \cite{Bak1} \cite{Bak2} may be converted into
a deterministic dynamical system by defining 
\begin{equation}
x_{i}\left( t+1\right) =\Gamma _{i}\left( \underset{\sim }{x}\right)
x_{i}\left( t\right) +\left( 1-\Gamma _{i}\left( \underset{\sim }{x}%
\right) \right) f\left( x_{i}\left( t\right) \right)   \label{2.1}
\end{equation}
where $\underset{\sim }{x}=\left\{ x_{i}\right\} $ is the vector of agent
coordinates and $f$ a deterministic pseudo-random generator, for example 
\[
f\left( x_{i}\right) =kx\hspace{1cm}\textnormal{mod}.1
\]
$k=2,3,\cdots $ . $\Gamma _{i}\left( \underset{\sim }{x}\right) $ is a
function which is nearly zero if $i$ corresponds to the agent with the
minimum $x_{i}$ value or to one of its $2n_{v}$ neighbors and is nearly one
otherwise. In Ref.\cite{Vilela1} the following function was proposed%
\footnote{%
Notice that in Ref.\cite{Vilela1} there is a typing mistake in the
definition of this function} 
\begin{equation}
\Gamma _{i}^{(1)}\left( \underset{\sim }{x}\right)
=\prod_{j=i-n_{V}}^{j=i+n_{V}}\left( 1-\prod_{l\neq j}\left( 1+e^{-\alpha
\left( x_{l}-x_{j}\right) }\right) ^{-1}\right)   \label{2.2}
\end{equation}
which, for large $\alpha $, satisfies the above conditions.

Here, instead, the function 
\begin{equation}
\Gamma _{i}^{(2)}\left( \underset{\sim }{x}\right)
=\prod_{j=i-n_{V}}^{j=i+n_{V}}\left( 1-\frac{e^{-x_{j}/T}}{%
\sum_{l=1}^{N}e^{-x_{l}/T}}\right)  \label{2.3}
\end{equation}
will be used, which has a similar behavior for $T\rightarrow 0^{+}$. Using a
similar function in a stochastic model Head\cite{Head} has shown the
breakdown of scale-free behavior for finite non-zero $T$. By using $\Gamma
_{i}^{(2)}\left( \underset{\sim }{x}\right) $ in the deterministic model
one may compare the Lyapunov spectrum analysis with the numerical results of
Head. In the $T\rightarrow 0$ limit both models are equivalent to the
original Bak-Sneppen model.

\begin{figure}[htb]
\begin{center}
\psfig{figure=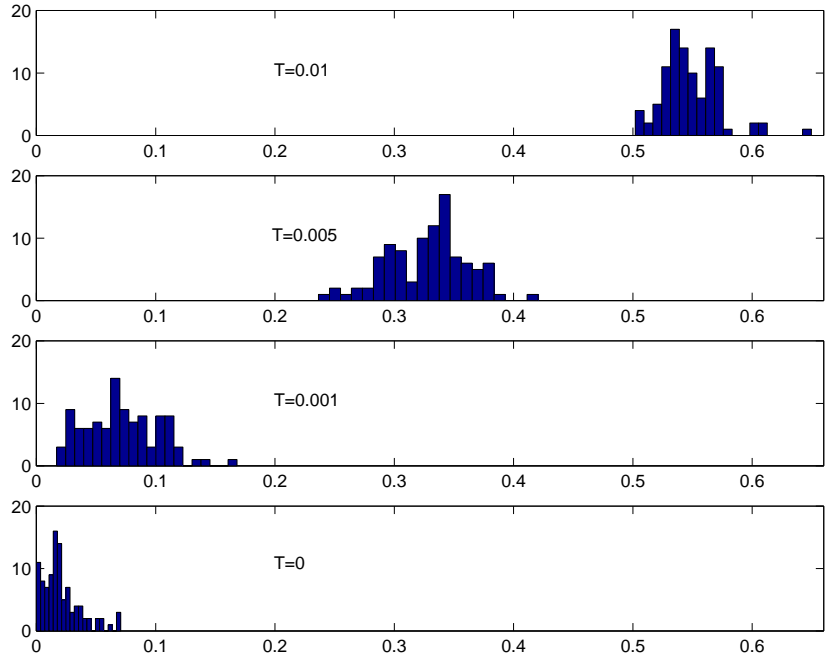,width=11truecm}
\end{center}
\caption{Approximate Lyapunov spectrum for several $T$ values in the
deterministic Bak-Sneppen model}
\end{figure}

Fig.1 shows an approximate calculation of the evolution of the Lyapunov
spectrum for successively decreasing $T$ values with $N$ agents. The results
were obtained for $N=100$, $1000$ time steps and $2n_{V}=2$. The spread in
the $T=0$ limit, around the exact value for $N$ agents $\left( \frac{3}{N}%
\log 2=0.02\text{ in this case}\right) $, results from the finite number of
time steps. The deterministic system defined by (\ref{2.1})-(\ref{2.3})
closely corresponds to the stochastic system studied in Ref.\cite{Head}.
Therefore the behavior of the Lyapunov spectrum explains why the
distribution of jump sizes and activation times obeys power laws only in the 
$T\rightarrow 0$ limit. In Ref. \cite{Head} the loss of criticality for
finite $T$ is interpreted in terms of the correlated or uncorrelated nature
of the active site jumps. The Lyapunov spectrum interpretation is clearer.
Notice also that it is only in the $N\rightarrow \infty $ limit that all the
Lyapunov exponents reach $0^{+}$. It is only in this limit that all time
scales disappear \cite{Vilela1}.

One of the most remarkable features of the Bak-Sneppen model is the
existence of a sharp probability distribution for the agents coordinates
above a barrier $b\simeq 0.667$. However, this sharp (full-measure)
distribution is merely a fixed point for the one-agent marginal, which is
the one-dimensional projection of a zero-measure subset of the invariant
measure in the $N-$ dimensional space \cite{Vilela1}. The fact that in the $%
N-$dimensional space the self-organized set to which the system returns
after each avalanche is of zero measure is consistent with Kac's lemma.
Kac's lemma states that for an ergodic invariant measure $\mu $ the average
return time to a set $A$ of measure $\mu \left( A\right) $ is $1/\mu \left(
A\right) $. Therefore for a scaling law $\rho \left( k\right) \sim 1/k^{\tau
}$, $\tau \leq 2$ implies $\mu \left( A\right) =0$.

To exhibit the zero-measure nature of the Bak-Sneppen state consider the 
{\it distance process} $d$ defined by 
\begin{equation}
d=\sum_{i}\max \left( b-x_{i},0\right)  \label{2.4}
\end{equation}

\begin{figure}[htb]
\begin{center}
\psfig{figure=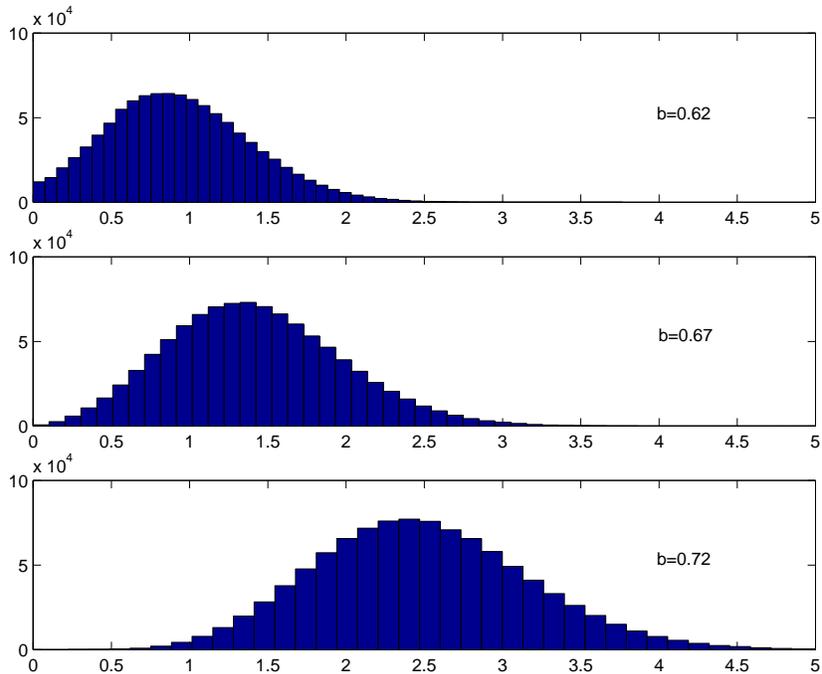,width=11truecm}
\end{center}
\caption{Probability density for the distance process}
\end{figure}

Fig.2 shows the probability distribution of the distance process for several
values of the barrier $b$. One sees that it is for $b$ around $0.67$ that
the neighborhood of the $d=0$ point becomes of zero measure.

The Bak-Sneppen self-organized state is an $N-$dimensional hypercube of
volume $\left( 1-0.667\right) ^{N}$. This set has repelling directions
corresponding to the agents that are active and neutral directions for all
others. Not being an invariant set it falls outside the usual definition of
``weak repeller''. It has been called a ``ghost weak repeller'' in Ref. \cite
{Vilela1}.

As mentioned before, the zero measure of this ``repeller'' makes the direct
measurement of the distribution law of avalanches a delicate matter. On the
one hand the barrier value defining the avalanches must be placed close to
the critical barrier to avoid the exponential finite-measure effects.
However because the average size of the avalanches is $1/\mu \left( A\right) 
$, the closer we are to the critical barrier the worse the statistics
becomes. A bin size has to be chosen to fit the avalanches data to a power
law and the accuracy of the exponent is found to be sensitive to the bin
size.

Logarithmic binning may improve the situation. However a more robust method
is based on the construction of the characteristic function 
\begin{equation}
C\left( x\right) =\left\langle e^{ikx}\right\rangle  \label{2.4a}
\end{equation}
from the data. For each avalanche size $k$, the probability density may then
be obtained by numerical computation of the inverse Fourier transform 
\begin{equation}
p\left( k\right) =\frac{1}{2\pi }\int_{-\pi }^{\pi }C\left( x\right)
e^{-ikx}dx  \label{2.4b}
\end{equation}
In this way all the data is used for each $k$ instead of only the events in
the neighborhood of $k$.

For sparse data this is a very robust way to construct $p\left( k\right) $,
which is no longer affected by the bin size choice. However in our case we
are not only interested in an accurate determination of $p\left( k\right) $
but also on the extraction of the power law prefactor that multiplies the
measure-dependent exponential. In cases where the exponent of the
exponential is also a non-trivial function of the measure it is extremely
difficult to separate the effect of the two terms. This is especially true
if the measure-dependent exponent decreases with the measure (see the
example below).

More accurate statements, concerning the nature of the prefactors, are
obtained by comparing the characteristic function constructed from the data
with the characteristic functions for trial distributions 
\begin{equation}
p\left( k,\mu \right) =ck^{-\alpha \left( \mu \right) }e^{-\nu _{\alpha
}\left( \mu \right) k}  \label{2.5}
\end{equation}
with $c$ and $\nu _{\alpha }\left( \mu \right) $obtained from normalization
and Kac's lemma, $\left\langle k\right\rangle =\frac{1}{\mu }$ . This was
done for several values of the measure, the results being displayed in Fig.3
(for $2n_{v}=2$). Both the exponential factor $\nu $ and the scaling
exponent $\alpha $ depend on the measure of the avalanche reference set.

\begin{figure}[htb]
\begin{center}
\psfig{figure=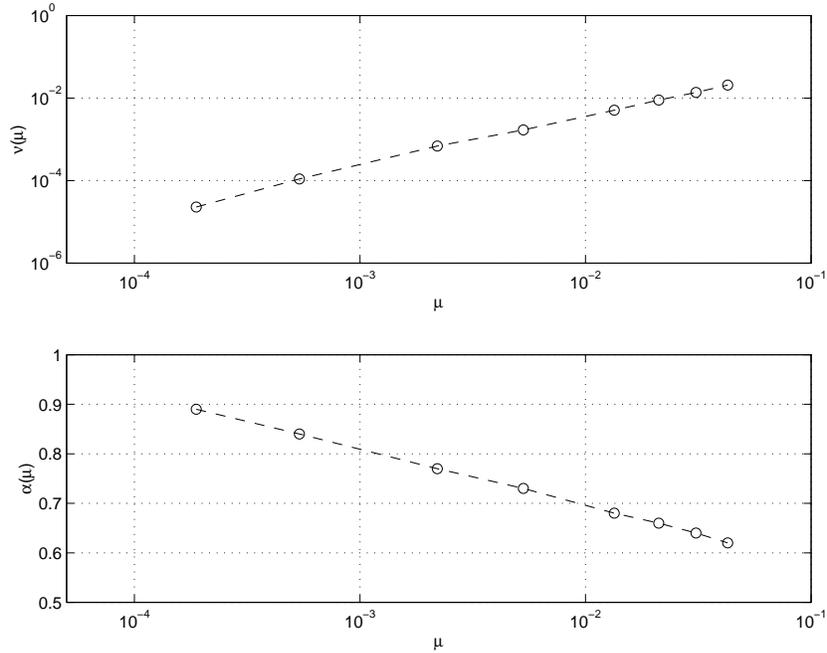,width=11truecm}
\end{center}
\caption{The exponential argument $\nu (\mu )$ and scaling exponent $\alpha
(\mu )$}
\end{figure}

This puts into evidence the unavoidable uncertainties in the direct
evaluation of the scaling exponent, in particular because even very close to
the critical barrier the scaling exponent still shows a noticeable
dependence on the measure of the reference set. In the Appendix the
characteristic functions associated to the probability distributions in Eq.(%
\ref{2.5}) are discussed in detail as well as some other results concerning
return times to small measure sets.

The discussion above refers to the problem of direct determination of the
scaling exponents. However, with the additional assumption, as done by
several authors, of a scaling form for $p\left( k\right) $ near the critical
barrier, an estimate of the $\mu \rightarrow 0$ value may be obtained. This
however relies on the validity of the scaling assumption. Assuming that
close to $\mu =0$%
\begin{equation}
p\left( k,\mu \right) =k^{-\alpha }f\left( k^{s}\mu \right)   \label{2.6}
\end{equation}
one obtains 
\begin{equation}
\begin{array}{lll}
\left\langle k\right\rangle  & \sim  & \mu ^{\frac{\alpha -2}{s}} \\ 
\left\langle k^{2}\right\rangle  & \sim  & \mu ^{\frac{\alpha -3}{s}}
\end{array}
\label{2.7}
\end{equation}
Then, from Kac's lemma, one obtains 
\[
s=2-\alpha 
\]
and from the numerical data for $2n_{v}=2$ (Fig. 4a), $\frac{\alpha -3}{s}%
\simeq 2.07$, leading to $\alpha \simeq 1.067$, $s\simeq 0.93$.

\begin{figure}[htb]
\begin{center}
\psfig{figure=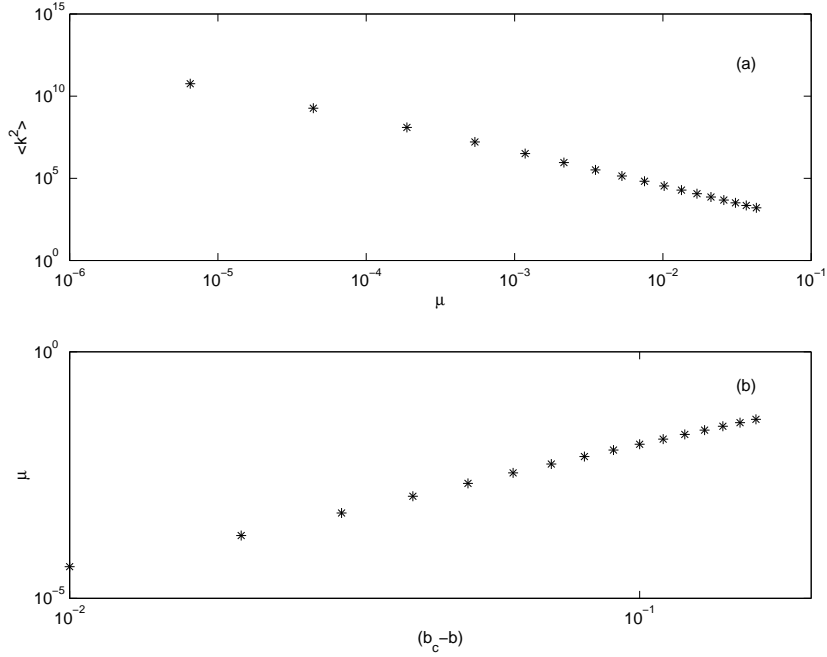,width=11truecm}
\end{center}
\caption{Data points used to infer the exponents in $\left\langle
k^{2}\right\rangle \sim \mu ^{\frac{\alpha -3}{s}}$ and $\mu \sim \left(
b_{c}-b\right) ^{\eta }$}
\end{figure}

Another exponent relates the measure to the position of the barrier 
\[
\mu \sim \left( b_{c}-b\right) ^{\eta } 
\]
with (Fig. 4b) $\eta \simeq 2.55$.

\section{The avalanche equation}

Assuming a differentiable dependence of the avalanches on the position of
the barrier, Maslov \cite{Maslov} derived an equation for the probability
density of avalanche sizes in the Bak-Sneppen model. Here, a similar
equation will be obtained for the return times to a measure $\mu $ set. This
equation is then applied to the Bak-Sneppen model. Particular attention is
paid to the avalanche merging kernel, which in the Maslov equation is
usually assumed to scale like a power in the critical region, but that may
actually have a much more complex form.

Assume that a set parametrization may be chosen in such a way that set
measures change in a differentiable manner. Differentiating with respect to
the measure, one obtains the following equation 
\begin{equation}
dp\left( k\right) =p\left( k\right) \left( 2\gamma \left( k,\mu \right) d\mu
-\frac{d\mu }{\mu }\right) -\sum_{j=1}^{k-1}p\left( j\right) \gamma \left(
j,\mu \right) d\mu p\left( k-j\mid j\right)  \label{3.1}
\end{equation}
where 
\[
P\left( T^{A\rightarrow A}=k,\mu \left( A\right) \right) \doteq p\left(
k\right) 
\]
Eq.(\ref{3.1}) is derived by shrinking the measure of the set $A$ by the
amount $d\mu $ and accounting for the number of avalanches of size $k$ that
become avalanches of larger size and for the number of mergers of smaller
avalanches leading to new $k-$avalanches. $\gamma \left( k,\mu \right) d\mu $
is the probability that the first or the last step of a $k-$avalanche
iteration falls in $d\mu $. The first term in Eq.(\ref{3.1}) accounts both
for a $k-$avalanche becoming a larger avalanche and for the change in the
total number of events from $N$ to $N\left( 1+\frac{d\mu }{\mu }\right) $.
The second term accounts for the probability for a $j-$avalanche and a $%
\left( k-j\right) -$avalanche to merge into a $k-$avalanche. $p\left(
k-j\mid j\right) $ is the conditional probability to have a $\left(
k-j\right) $ avalanche following a $j$ one.

With the additional hypothesis of statistical independence between
successive avalanches 
\begin{equation}
p\left( k-j\mid j\right) =p\left( k-j\right)  \label{3.2}
\end{equation}

Defining $C\left( x,\mu \right) =\left\langle \exp \left( ikx\right)
\right\rangle $ (the characteristic function) and $Q\left( x,\mu \right)
=\left\langle \gamma \left( k,\mu \right) \exp \left( ikx\right)
\right\rangle $ one obtains 
\begin{equation}
\frac{d}{d\mu }C\left( x,\mu \right) =2Q\left( x,\mu \right) -\frac{1}{\mu }%
C\left( x,\mu \right) -Q\left( x,\mu \right) C\left( x,\mu \right)
\label{3.3}
\end{equation}

So far this is a very general equation that (with statistical independence
of avalanches) would apply to any dynamical system. The dynamical
information of each system is coded on the ``merging kernel'' $\gamma (k,\mu
)$. Using normalization $\left( C\left( 0,\mu \right) =1\right) $ and Kac's
lemma $\left( -i\frac{d}{dx}C\left( x,\mu \right) \mid _{x=0}=\frac{1}{\mu }%
\right) $ in Eq.(\ref{3.3}) one obtains the following constraints on $\gamma
(k,\mu )$%
\begin{equation}
\begin{array}{lll}
\left\langle \gamma (k,\mu )\right\rangle  & = & \frac{1}{\mu } \\ 
\left\langle k\gamma (k,\mu )\right\rangle  & = & \frac{1}{\mu ^{2}}
\end{array}
\label{3.4a}
\end{equation}
A trivial solution to the constraints (\ref{3.4a}) is 
\begin{equation}
\gamma (k,\mu )=\frac{1}{\mu }  \label{3.5}
\end{equation}
leading to the equation 
\begin{equation}
\mu \frac{dC\left( x,\mu \right) }{d\mu }=C\left( x,\mu \right) -C^{2}\left(
x,\mu \right)   \label{3.6}
\end{equation}
with solution 
\begin{equation}
C\left( x,\mu \right) =\frac{\mu }{\mu +q\left( x\right) }  \label{3.7}
\end{equation}
with $q\left( 0\right) =0$ and $q^{^{\prime }}\left( 0\right) =-i$, from
normalization and Kac's lemma. Otherwise $q\left( x\right) $ is an arbitrary
function of $x$.

\begin{figure}[htb]
\begin{center}
\psfig{figure=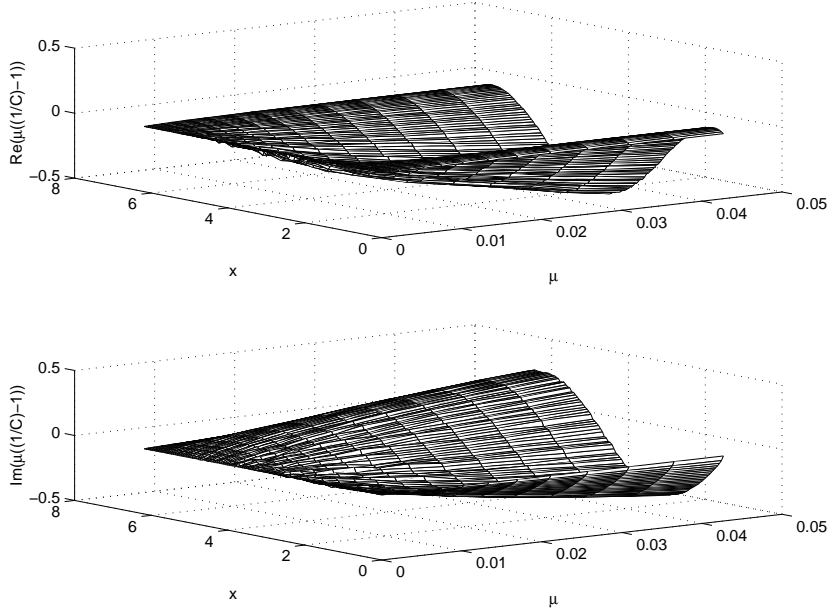,width=11truecm}
\end{center}
\caption{Real and imaginary parts of $\mu \left( \frac{1}{C(x,\mu )}-1\right) 
$}
\end{figure}

However the Bak-Sneppen model is not in the class of solutions (\ref{3.7}),
because from this equation it follows that $\mu \left( \frac{1}{C\left(
x,\mu \right) }-1\right) $ should be independent of $\mu $, which is not
verified by the numerical data (Fig. 5).

A power dependence of the merger kernel $\gamma (k,\mu )$%
\[
\gamma (k,\mu )=\alpha _{1}\frac{1}{\mu }+\alpha _{2}k^{\varepsilon } 
\]
together with the constraints (\ref{3.4a}) implies 
\[
\left\langle k^{\varepsilon }\right\rangle _{\mu }=\mu \left\langle
k^{\varepsilon +1}\right\rangle _{\mu } 
\]
which again, is not borne out by the data.

However from the numerical simulation data ($2n_{v}=2$) for two close values
of the measure and Eq.(\ref{3.3}) one may obtain direct information on $%
\gamma (k,\mu )$ from 
\[
\gamma (k,\mu )=\frac{\int_{-\pi }^{\pi }Q\left( x,\mu \right) \exp \left(
-ikx\right) dx}{\int_{-\pi }^{\pi }C\left( x,\mu \right) \exp \left(
-ikx\right) dx}
\]

\begin{figure}[htb]
\begin{center}
\psfig{figure=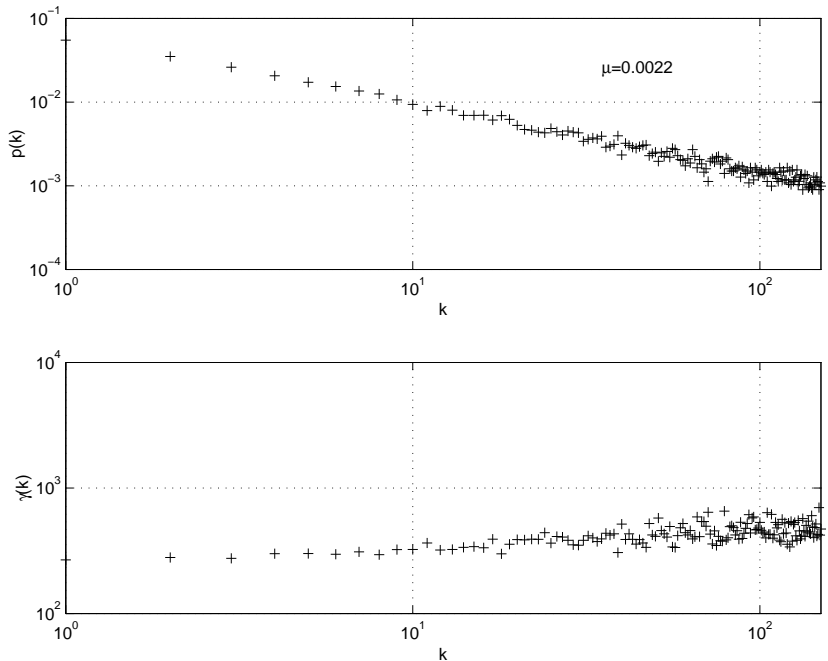,width=11truecm}
\end{center}
\caption{$p(k)$ and $\gamma (k)$ for $\mu =0.0022$}
\end{figure}

An example is shown in Fig. 6, where both $p\left( k,\mu \right) $ and $%
\gamma \left( k,\mu \right) $ are plotted. The general conclusion from these
computations for several measures is that, starting from a non-zero value at 
$k=1$, $\gamma \left( k,\mu \right) $ grows relatively fast to the
asymptotic value $\frac{1}{\mu }$. A (weak) power dependence is only found
for small $k$ values. Therefore the behavior of the merger kernel, that
codes the dynamical information of the Bak-Sneppen model, seems to be much
more complex than thought before.

\section{Conclusions}

(i) The deterministic Bak-Sneppen model, studied in this paper, establishes
a clear relation between the Lyapunov spectrum and scale-free behavior.

Extremal behavior, where only a finite number of agents is active at each
time, leads to the vanishing of all Lyapunov exponents in the infinite
agents limit. On the other hand, local chaos insures that the convergence to
zero is from above. These seem to be the most common ingredients in systems
that display what has been called self-organized criticality.

(ii) The self-organized state in the deterministic Bak-Sneppen model is an
interesting zero-measure subset of the invariant measure. Interpreting
avalanches as return times to a set $A$, one concludes that a scaling law
with exponent $\leq 2$ implies, by Kac's lemma, $\mu \left( A\right) =0$.
Therefore one expects the zero measure feature to be present not only in the
Bak-Sneppen model but also in other self-organized states with avalanche
scaling laws.

(iii) The avalanche equation, in the measure formulation discussed in
Section 3, is a powerful guide to constraint the extraction of dynamical
information from dynamical systems of this type. In particular it provides
preliminary evidence on a non-power behavior of the merger kernel. A
stretched exponential behavior converging to $\frac{1}{\mu }$ seems to be a
better guess. This however is a question that deserves further study.

\section{Appendix. Return times to small measure sets}

On a measurable space $\left( M,{\cal B},\mu \right) $, consider a
differentiable map $T$ for which $\mu $ is an ergodic invariant measure. The 
{\it return time} $\tau _{A}\left( x\right) $ to a measurable set $A$
starting from a point $x\in M$ is 
\begin{equation}
\tau _{A}\left( x\right) =\inf \left\{ \left( k\geq 1\mid T^{k}x\in A\right)
\cup \infty \right\}  \label{A1}
\end{equation}
The probability to find a return time greater than $k$, starting from an
arbitrary point $x\in M$ 
\begin{equation}
P\left( T^{M\rightarrow A}>k\right) =\mu \left( \tau _{A}\left( x\right)
>k\right)  \label{A2}
\end{equation}
and for a starting point $x$ in $A$%
\begin{equation}
P\left( T^{A\rightarrow A}>k\right) =\mu _{A}\left( \tau _{A}\left( x\right)
>k\right)  \label{A3}
\end{equation}
where 
\begin{equation}
\mu _{A}\left( \tau _{A}\left( x\right) >k\right) =\frac{\mu \left( A\cap
\left\{ \tau _{A}\left( x\right) >k\right\} \right) }{\mu \left( A\right) }
\label{A4}
\end{equation}
Invariance of the measure implies 
\begin{eqnarray}
\mu \left( \tau _{A}\left( x\right) >k\right) &=&\mu \left( A^{c}\cap
\left\{ \tau _{A}\left( x\right) >k-1\right\} \right)  \label{A5} \\
&=&\mu \left( \tau _{A}\left( x\right) >k-1\right) -\mu \left( A\right) \mu
_{A}\left( \tau _{A}\left( x\right) >k-1\right)  \nonumber
\end{eqnarray}
and because $P\left( T^{M\rightarrow A}=k\right) =\mu \left( \tau _{A}\left(
x\right) >k-1\right) -\mu \left( \tau _{A}\left( x\right) >k\right) $%
\begin{equation}
P\left( T^{M\rightarrow A}=k\right) =\mu \left( A\right) P\left(
T^{A\rightarrow A}>k-1\right)  \label{A6}
\end{equation}
Because of ergodicity, Poincar\'{e} recurrence implies $\sum_{k=1}^{\infty
}P\left( T^{M\rightarrow A}=k\right) =1$. On the other hand $%
\sum_{k=1}^{\infty }P\left( T^{A\rightarrow A}>k-1\right)
=\sum_{k=1}^{\infty }kP\left( T^{A\rightarrow A}=k\right) $ and for $\mu
\left( A\right) \neq 0$ one obtains Kac's lemma 
\begin{equation}
{\Bbb E}\left( \tau _{A}\left( x\right) \mid x\in A\right) =\int_{A}\tau
_{A}\left( x\right) d\mu _{A}\left( x\right) =\frac{1}{\mu \left( A\right) }
\label{A7}
\end{equation}

Defining, as in \cite{Sandro1} 
\begin{equation}
c\left( k,A\right) =\mu _{A}\left( \tau _{A}\left( x\right) >k\right) -\mu
\left( \tau _{A}\left( x\right) >k\right)  \label{A8}
\end{equation}
one obtains from (\ref{A5}) and (\ref{A3}) 
\[
P\left( T^{A\rightarrow A}>k\right) =\left( 1-\mu \left( A\right) \right)
P\left( T^{A\rightarrow A}>k-1\right) -\Delta c\left( k,A\right) 
\]
where $\Delta c\left( k,A\right) =c\left( k-1,A\right) -c\left( k,A\right) $%
. By iteration 
\begin{equation}
P\left( T^{A\rightarrow A}>k\right) =\left( 1-\mu \left( A\right) \right)
^{k}-\sum_{j=1}^{k}\Delta c\left( j,A\right) \left( 1-\mu \left( A\right)
\right) ^{k-j}  \label{A9}
\end{equation}
Let 
\begin{equation}
\Delta c\left( A\right) =\sup_{k}\left| \Delta c\left( k,A\right) \right|
\label{A10}
\end{equation}
Then, by standard techniques, as in \cite{Sandro1}, one obtains the
estimation 
\begin{equation}
\left| P\left( T^{A\rightarrow A}>k\right) -e^{-k\mu \left( A\right)
}\right| \leq 2\mu \left( A\right) +\Delta c\left( A\right) \left( 1-\log
\Delta c\left( A\right) \right)  \label{A11}
\end{equation}
This estimate is sharper than the one obtained in \cite{Sandro1} because it
involves $\Delta c\left( A\right) $ rather than $c\left( A\right)
=\sup_{k}\left| c\left( k,A\right) \right| $.

The estimate (\ref{A11}) holds whenever $\mu \left( A\right) \neq 0$.
However it is useless to find $p(k,\mu )=P\left( T^{A\rightarrow A}=k\right) 
$ in the $\mu \left( A\right) \rightarrow 0$ limit because 
\[
e^{-\left( k-1\right) \mu \left( A\right) }-e^{-k\mu \left( A\right)
}=e^{-k\mu \left( A\right) }\left( e^{-\mu \left( A\right) }-1\right) 
\underset{\mu \left( A\right) \rightarrow 0}{\rightarrow }0 
\]
Therefore, a different approach must be followed to control the $\mu \left(
A\right) \rightarrow 0$ limit.

For general distributions of the form 
\[
p_{g}\left( k,\mu \right) \sim g\left( k\right) e^{-k\nu \left( \mu \right)
} 
\]
the characteristic function must satisfy the equation 
\[
\frac{d}{d\nu }C\left( x,\mu \right) =i\frac{d}{dx}C\left( x,\mu \right) +%
\frac{1}{\mu }C\left( x,\mu \right) 
\]
with solution 
\[
C\left( x,\mu \right) =\exp \left( \int_{\nu _{0}}^{\nu }\frac{d\tau }{\mu
\left( \tau \right) }-\int_{\nu _{0}}^{\nu -ix}\frac{d\tau }{\mu \left( \tau
\right) }\right) 
\]
Therefore a large family of solutions is obtained simply by specifying the
function $\nu \left( \mu \right) $ (and its inverse $\mu \left( \nu \right) $%
).

A particular case is 
\[
p_{\alpha }\left( k,\mu \right) \sim k^{-\alpha }e^{-k\nu _{\alpha }\left(
\mu \right) } 
\]
For $\alpha =0$ this is the geometric distribution with 
\[
\nu _{\alpha }\left( \mu \right) =-\log \left( 1-\mu \right) 
\]
and for $\alpha =1$%
\[
C_{1}\left( x,\mu \right) =\frac{\log \left( 1-\exp \left( ix-\nu \right)
\right) }{\log \left( 1-\exp \left( -\nu \right) \right) } 
\]
with 
\[
\mu =\left( 1-e^{\nu }\right) \log \left( 1-e^{-\nu }\right) 
\]
The functions $\nu _{\alpha }\left( \mu \right) $ for other values of $%
\alpha $ are plotted in Fig.7.

\begin{figure}[htb]
\begin{center}
\psfig{figure=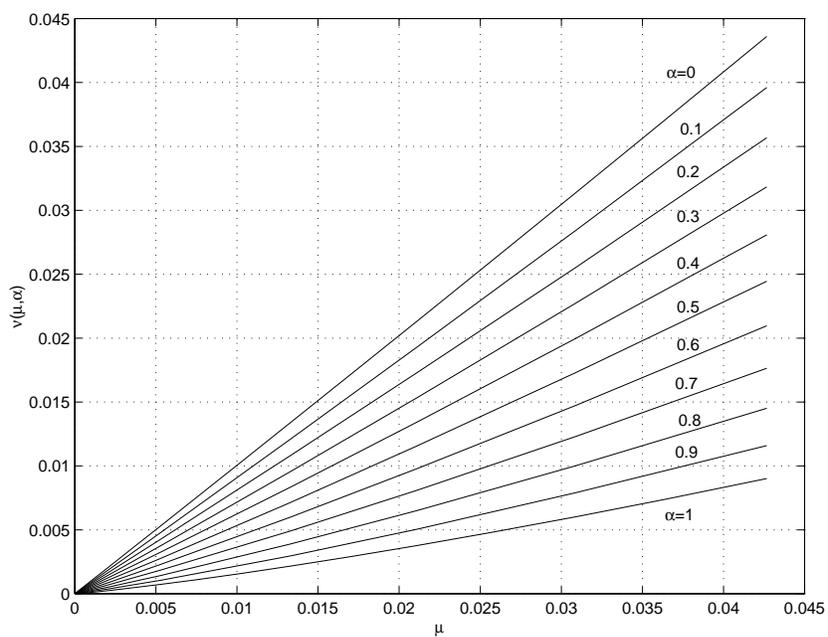,width=11truecm}
\end{center}
\caption{Exponential argument $\nu _{\alpha }(\mu )$ for the probability
densities $p_{\alpha }\left( k,\mu \right) \sim k^{-\alpha (\mu )}\exp
\left( -k\nu _{\alpha }(\mu )\right) $}
\end{figure}

\end{document}